%% file: main.tex
\documentclass[10pt,conference]{IEEEtran}
\usepackage{cite}
\usepackage{amsmath,amssymb,amsfonts}
\usepackage{mathdots}
\usepackage{algorithmicx}
\usepackage{graphicx}
\usepackage{caption}
\usepackage{subcaption}
\usepackage{textcomp}
\usepackage[dvipsnames]{xcolor}
\usepackage{orcidlink}
\usepackage{hyperref}
\usepackage{todonotes}
\usepackage{pgfplots}
\usepackage{pgf}
\usepackage{svg}
\usepackage{multirow}
\usepackage{bbm}

\usepackage{booktabs}
\usepackage{siunitx}

\usepackage[english]{babel}
\usepackage{amsthm}
\theoremstyle{definition}

\usepackage{algorithm2e}
\RestyleAlgo{ruled}
\usepackage[noend]{algpseudocode}
\usepackage{dsfont}
\usepackage{listings}

\usepackage[capitalise]{cleveref}

\usepackage{listings}
\lstdefinelanguage{json}{
  basicstyle=\ttfamily\small,
  showstringspaces=false,
  breaklines=true,
  frame=single
}
\lstdefinelanguage{rsqasm}{
  morekeywords={RSQASM, cz, h, s, t, rx, ry, rz, move},
  basicstyle=\ttfamily\small,
  showstringspaces=false,
  breaklines=true,
  frame=single
}

\usepackage[capitalise]{cleveref}

\def\BibTeX{{\rm B\kern-.05em{\sc i\kern-.025em b}\kern-.08em
    T\kern-.1667em\lower.7ex\hbox{E}\kern-.125emX}}    
\begin{document}
\title{Practical Insights into Fair Comparison and Evaluation Frame for Neutral-Atom Compilers\\
{
}
\thanks{The research is funded by the Munich Quantum Valley (MQV), which is supported by the Bavarian state government with funds from the Hightech Agenda Bayern Plus, and by the German Federal Ministry of Research, Technology and Space (BMFTR) under the funding program Quantum Technologies - From Basic Research to Market under contract number 13N16087. We are also grateful to the \textit{
Quantum Compiler team} at the Leibniz Supercomputing Centre (LRZ) for many fruitful discussions.}
}

\author{
\IEEEauthorblockN{
Emil Khusainov \orcidlink{0009-0000-6605-0965},
Yanbin Chen \textsuperscript{1,2}\orcidlink{0000-0002-1123-1432},
Jonas Winklmann \textsuperscript{3}\orcidlink{0009-0009-4108-7732},
Helmut Seidl \textsuperscript{1}\orcidlink{0000-0002-2135-1593},
Christian B. Mendl \textsuperscript{2}\orcidlink{0000-0002-6386-0230}
}

\IEEEauthorblockA{\textsuperscript{1}
Chair of Programming Languages, Compiler Construction and Specification Formalisms}

\IEEEauthorblockA{\textsuperscript{2}
Chair of Scientific Computing}

\IEEEauthorblockA{\textsuperscript{3}
Chair of Computer Architecture and Parallel Systems}

\IEEEauthorblockA{
\textit{TUM School of Computation, Information and Technology} \\
\textit{Technical University of Munich}, Germany \\
{emil.khusainov, yanbin.chen, jonas.winklmann, helmut.seidl, christian.mendl}@tum.de
}
}

\maketitle

\begin{abstract}
\input{sections/abstract}
\end{abstract}


\section{Introduction}\label{sec:intro}
\input{sections/introduction}

\section{Preliminaries}\label{sec:prelim}
\input{sections/preliminaries}

\section{Method}\label{sec:methods}
\input{sections/method}

\section{Evaluation}\label{sec:evaluation}
\input{sections/evaluation}

\section{Related work}\label{sec:rel_works}
\input{sections/relevant_work}

\section{Conclusion}\label{sec:conclusion}
\input{sections/conclusion}

\section*{Acknowledgment}
The research is funded by the Munich Quantum Valley (MQV), which is supported by the Bavarian state government with funds from the Hightech Agenda Bayern Plus, and by the German Federal Ministry of Research, Technology and Space (BMFTR) under the funding program Quantum Technologies - From Basic Research to Market under contract number 13N16087. We are also grateful to the \textit{
Quantum Compiler team} at the Leibniz Supercomputing Centre (LRZ) for many fruitful discussions.
\bibliographystyle{IEEEtran}
\bibliography{references}

\end{document}

%% file: sections/abstract.tex
Neutral-atom quantum computing is among the most promising platforms for scalable quantum computation, and compilation toolchains are crucial for leveraging capabilities such as qubit shuttling and parallel gate execution. An important challenge, however, is that existing neutral-atom compilers are often evaluated using metrics computed over different parts of the toolchain and under non-equivalent assumptions. Consequently, fair quantification and comparison of compiler performance remain difficult. Reported metrics may depend on inconsistent transpilation optimization levels, different movement-duration models, different sets of considered fidelity sources, and even minor implementation bugs or undocumented representation choices.
To address this problem, we present a unified and reproducible evaluation framework for neutral-atom compilers. Our framework introduces RSQASM (Routed and Scheduled QASM), a QASM-inspired post-compilation representation that captures mapped, routed, and scheduled circuits, including explicit parallel gate execution and shuttling operations.
As part of the framework, we provide adapter scripts that translate existing compiler outputs and intermediate artifacts into RSQASM.
As a case study, we compare three well-known neutral-atom compilation toolchains: HybridMapper, DasAtom, and Enola, motivated by the large performance differences reported in prior work. Using our framework and representation, we perform a new evaluation and show that several previously claimed performance gaps become substantially smaller and, in some cases, are not reproduced once evaluation inconsistencies are removed.

%% file: sections/introduction.tex
Neutral-atom quantum computing has emerged as one of the most promising directions toward scalable quantum computation. Its support for large atom arrays, qubit shuttling, and parallel gate execution makes it a natural target for specialized compilation toolchains \cite{Wintersperger_2023, Bluvstein2022-bf, Graham2022-pl, Barnes2022-gh, Henriet2020quantumcomputing}. Consequently, a growing number of neutral-atom compilers have been proposed to map, route, and schedule quantum circuits while exploiting these hardware-specific capabilities \cite{schmid2023hybridcircuitmappingleveraging, huang2025dasatomdivideandshuttleatomapproach, Enola, wang2024atomiquequantumcompilerreconfigurable}.

A persistent challenge in comparing these solutions is that existing compiler studies typically rely on toolchain-specific fidelity models and evaluation assumptions, which makes direct benchmark comparisons difficult. This broader problem has already been discussed in the quantum computing literature \cite{lall2025reviewcollectionmetricsbenchmarks}. Our focus, however, is not cross-hardware benchmarking but evaluation consistency across neutral-atom compilers.

In our previous hardware-agnostic benchmarking study \cite{Bachelorarbeit_Emil_Khusainov}, we observed that existing neutral-atom compilers are often evaluated over different parts of the toolchain and under non-equivalent assumptions, which makes direct performance comparisons unreliable. The study identified several sources of distortion, including inconsistent fidelity accounting, different coherence and shuttling-time models, partially mismatched hardware parameterizations, and minor implementation bugs that materially affect the reported metrics. After aligning these assumptions and correcting the discovered issues, the more than 400x fidelity advantage between DasAtom and Enola on QFT30 previously reported by \cite{huang2025dasatomdivideandshuttleatomapproach} was reduced to about 5.5x. These findings exposed a broader unresolved problem: neutral-atom compilation still lacks a common post-compilation representation and a controlled, reproducible evaluation methodology.

However, our previous study was primarily diagnostic: it showed that obtaining even a roughly fair comparison required substantial manual alignment of architectures, metrics, and tool outputs, together with non-trivial pre-processing, mid-processing, and post-processing steps. Even under this effort, small inaccuracies could still remain and materially affect the conclusions. This suggests that fair evaluation of neutral-atom compilers cannot rely on case-by-case manual alignment and ad hoc corrections alone. Instead, it requires a reusable post-compilation interface and a unified evaluation pipeline that make assumptions explicit, normalize compiler outputs systematically, and support controlled comparisons across toolchains. 

\begin{figure}[t]
    \centering
    \includegraphics[width=\linewidth]{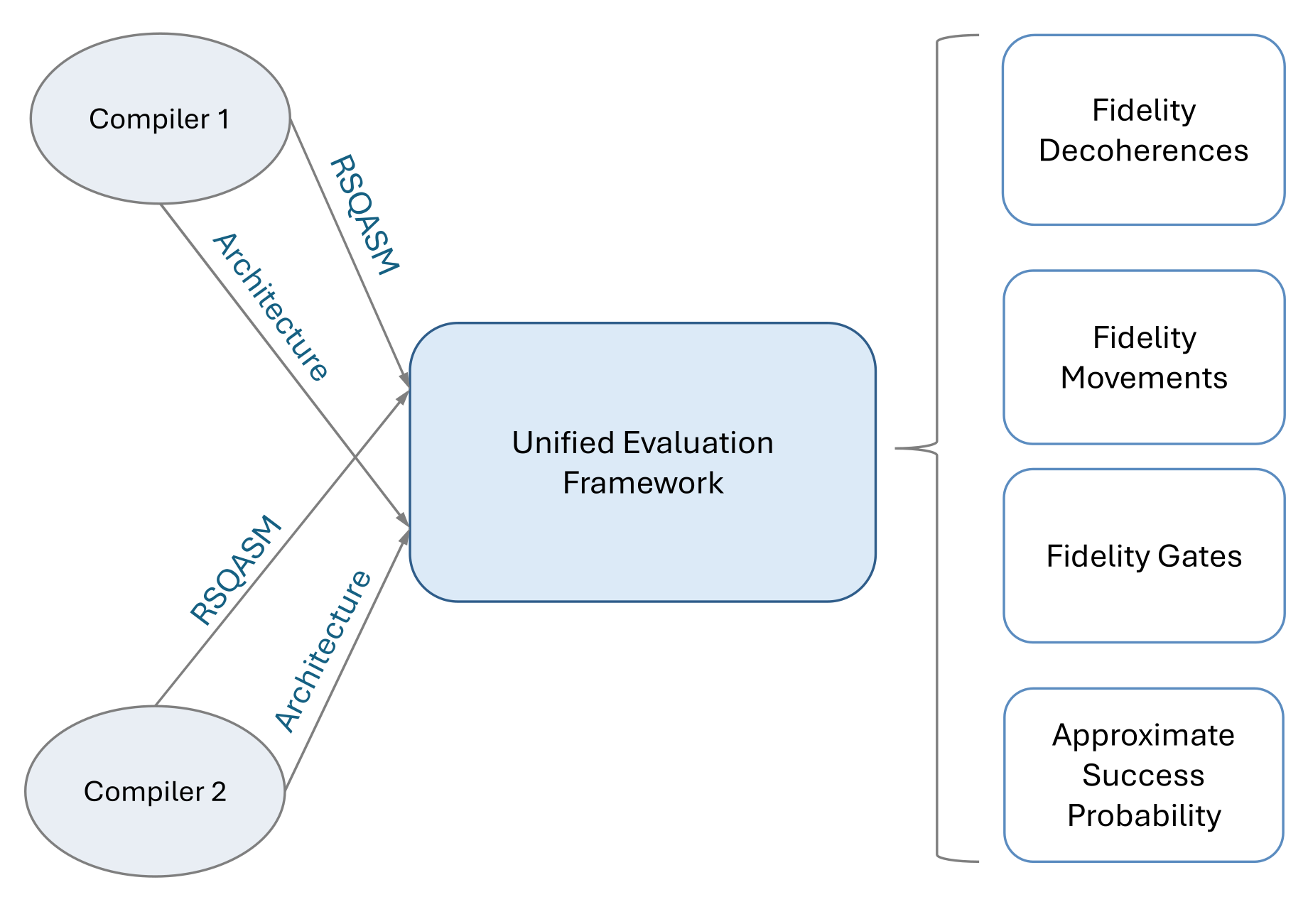}
    \caption{Overview of the proposed benchmarking framework. Different compilers produce compiled circuits and target hardware descriptions in the proposed format, abstracting away their internal representations.
    The unified evaluation module then processes this representation to compute fidelity-related metrics, as well as approximate success probability, enabling consistent and fair comparison.}
    \label{fig:concept-diagram_Abstract}
\end{figure}

To address this problem, we develop a unified evaluation framework that turns such ad hoc comparisons into a systematic and reusable process. As illustrated in \cref{fig:concept-diagram_Abstract}, the proposed workflow decouples compiler-specific outputs from the evaluation stage by translating them into a common post-compilation representation.

The evaluation stage applies a consistent set of metric definitions and hardware assumptions to compute fidelity-related quantities, including decoherences, movements, and gates fidelities, as well as the resulting approximate success probability. This design enables controlled, reproducible, and fair comparisons across different neutral-atom compilers.

Furthermore, the framework provides a practical interface for integrating future toolchains into the same evaluation pipeline. The main contributions of this work are summarized as follows:
\begin{itemize}
  \item We introduce a unified representation for mapped, routed, and scheduled neutral-atom quantum circuits, derived from the QASM syntax used in Qiskit
  \item We propose a benchmarking methodology based on relevant and observable sources of fidelity and success probability, inspired by existing distinct methodologies of considered prior compilers, as illustrative examples. In addition, we define a compact and easy-to-parse format for target hardware descriptions
  
  \item We demonstrate the applicability of the framework by extracting intermediate representations and outputs from selected compilers, translating them into the proposed format, and evaluating them under a unified methodology. The evaluation pipeline automates the core stages, enabling reproducible experiments and allowing users to evaluate arbitrary circuits across different compilers in a consistent manner. Furthermore, we identify and address implementation inconsistencies in existing compilers that could lead to inconsistent evaluation results, and incorporate these corrections into the evaluation workflow.
\end{itemize}

We expect the proposed framework to reduce inconsistencies in existing evaluation pipelines and to enable fairer comparisons across neutral-atom compiler toolchains. Moreover, the proposed format, RSQASM, captures the key post-compilation aspects of neutral-atom circuits and can serve as a common interface for future toolchains.

%% file: sections/preliminaries.tex
Neutral-atom quantum computing architectures can address qubit connectivity constraints not only through logical SWAP operations, but also through the physical movement of atoms, commonly referred to as shuttling. These architectures typically combine a Spatial Light Modulator (SLM) array for static trapping with Acousto-Optic Deflector (AOD) for dynamic atom transport. Furthermore, both gate execution and shuttling operations can be performed in parallel. We assume familiarity with the common physical and architectural constraints of neutral-atom quantum computing platforms \cite{Bluvstein2022-bf, Wintersperger_2023}.

To leverage these hardware capabilities, existing compilers adopt different mapping, routing, and scheduling strategies. In particular, routing strategies may rely on logical SWAP operations, physical qubit movements (shuttlings), or a combination of both (hybrid). These approaches are commonly referred to as SWAP-based, move-based, and hybrid routing, respectively. However, most of these compilers also rely on compiler-specific architectural abstractions and evaluation models, leading to substantial inconsistencies in reported benchmarking results. Consequently, fair comparison of compilation approaches remains difficult. In addition, tool-specific evaluation pipelines may obscure implementation bugs, as such issues can remain hidden under compiler-dependent assumptions and therefore compromise accurate cross-compiler assessment.

%% file: sections/method.tex
This section describes the proposed evaluation pipeline, the common representations used by the framework, the evaluator itself and adapter scripts for the selected compilers. The goal is to translate compiler-specific outputs into a shared post-compilation representation and then evaluate them under aligned hardware assumptions and metric definitions.

\subsection{Pipeline Overview}
\begin{figure}[t]
    \centering
    \includegraphics[width=\linewidth]{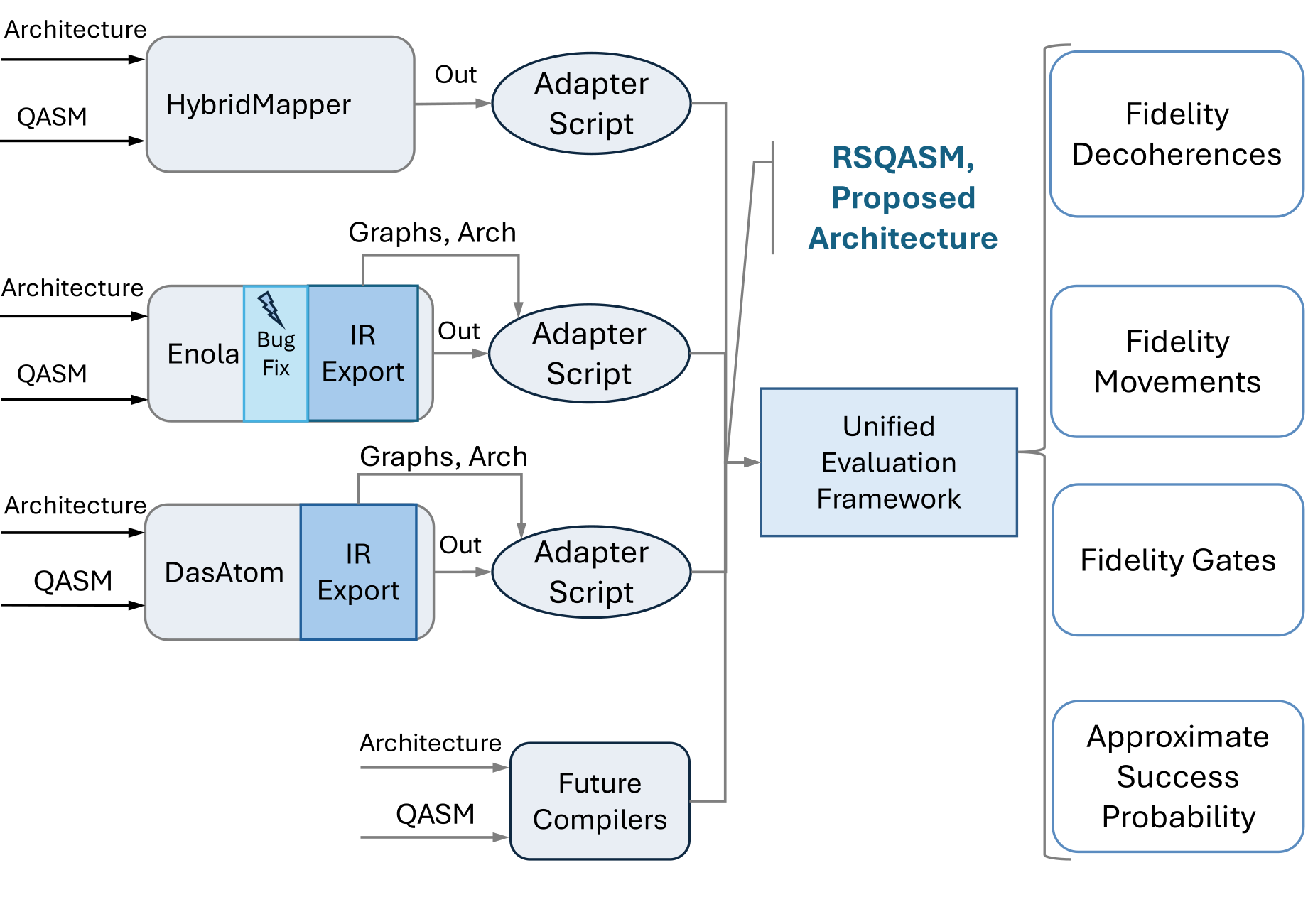}
    \caption{Detailed overview of the proposed benchmarking pipeline.
    The pipeline consists of three stages starting from left: extraction of compiled outputs and intermediate representations, translation into the proposed consistent representation, and evaluation of fidelity-related metrics under aligned assumptions.
    }
    \label{fig:concept-diagram_Detailed}
\end{figure}

\cref{fig:concept-diagram_Detailed} illustrates the overall workflow of the proposed benchmarking framework. The pipeline begins by running the three selected compilers on the same input QASM circuit together with their corresponding hardware descriptions. For Enola and DasAtom, the workflow additionally includes extraction of intermediate representations and, where necessary, correction of implementation issues to recover information corrupted in or omitted from the final compiler outputs.

In the second stage, the generated artifacts are passed to the corresponding adapter scripts, which translate them into the proposed target hardware architecture and RSQASM representation. The figure illustrates this intended integration model: existing compilers are connected through adapter scripts, whereas compiler toolchains that directly emit RSQASM and the proposed architecture description can be connected to the evaluator without an additional translation step.

Finally, the evaluator receives RSQASM and the target hardware architecture as inputs and simulates the execution process using an internal square hardware grid composed of discrete traps, whose parameters are defined by the architecture model. It manages qubit movements between cells, tracks cell occupancy, and accounts for travel distance and duration in order to compute decoherences fidelity, gates fidelity, and movements fidelity, as well as the resulting approximate success probability under a consistent set of assumptions.

In this representation, RSQASM abstracts away explicit qubit identifiers and instead refers to qubits by their cell coordinates, arranged in row-major order.

\subsection{Target Hardware Architecture Specification}
Because the selected compilers use different target hardware architecture descriptions, the framework requires a unified compact format that can be processed independently of any individual compiler. Inspired by the hardware description format used in HybridMapper \cite{schmid2023hybridcircuitmappingleveraging}, we propose a JSON-based architecture specification organized as follows:

\begin{itemize}
  \item \textbf{Root object}: \texttt{schema}, an integer specifying the schema version for future changes; \texttt{properties}, describing the square hardware grid of atom traps; and \texttt{parameters}, containing qubit-, gate- and circuit-level parameters.

  \item \textbf{properties}: \texttt{ nRows\_nColumns\_grid\_side\_size}, an integer specifying the number of cells along each side of the grid; \texttt{interQubitDistance}, the physical spacing between orthogonally adjacent qubits in \si{\micro\meter}. In the current work, only square grids are considered.
  
  \item \textbf{parameters}: \texttt{Qubits}, a list of working qubits on the grid; \texttt{gateTimes}, a specification of gate execution times in \si{\micro\second}; \texttt{gateFidelities}, a specification of gates fidelity parameters; \texttt{shuttlingTimesSpeed}, parameters related to shuttling duration; \texttt{shuttlingFidelities}, parameters related to shuttling-induced fidelity loss; and \texttt{decoherenceTimes}, specifies a set of \(T_1\) and \(T_2\) decoherence times in \si{\micro\second} used for decoherences fidelity calculation.
  
  \item \textbf{Qubits}: \texttt{id}, the qubit identifier; \texttt{x}, \texttt{y}, the initial physical coordinates, starting from top-left corner using screen-space coordinates.

  \item \textbf{shuttlingTimesSpeed}: \texttt{move\_speed}, shuttling speed in \si{\micro\meter\per\micro\second};
  \texttt{aod\_activate\_deactivate\_time}, the time associated with AOD activation and deactivation during shuttling.

  \item \textbf{shuttlingFidelities}: fidelity impact by transition between SLM and AOD via \texttt{aod\_activate\_deactivate}.
  
\end{itemize}

\subsection{RSQASM Structure}

RSQASM is the post-compilation representation used by the framework. It is based on QASM, but extends it with explicit shuttling operations and stage-based parallel execution. In addition, it abstracts away explicit qubit indices by representing qubits via their positions in the hardware grid. This design enables mapped, routed, and scheduled neutral-atom circuits to be represented in a compact, human-readable form together with the proposed hardware descriptions.

\begin{lstlisting}[language=rsqasm,
caption={Illustrative example of RSQASM format.},
label={lst:rsqasm}]
RSQASM 1.0;
h q[0];
cz q[2], q[1];
move q[3], q[4];
cz q[0], q[5];cz q[1], q[3];move q[2], q[4];
\end{lstlisting}
The first line declares the format identifier and schema version. 
Each line represents one execution stage, i.e., a set of operations applied in parallel to the qubits on the grid. For example, \texttt{h q[0];} applies a Hadamard gate to the qubit located at cell 0. Cell indices are enumerated in row-major order starting from the top-left corner. The instruction \texttt{cz q[2], q[1];} applies a controlled-Z gate to the qubits at cells 2 and 1. The instruction \texttt{move q[3], q[4];} moves a qubit from cell 3 to cell 4, provided that the destination cell is unoccupied. Finally, the line \texttt{cz q[0], q[5];cz q[1], q[3];move q[2], q[4];} represents a stage in which both CZ gates and one movement are executed simultaneously. In general, any combination of mutually compatible operations may appear within the same stage.

\subsection{Evaluator}
The purpose of the evaluator is to compare compiled neutral-atom circuits under a single, explicit, and reproducible set of assumptions. Existing compilers commonly rely on their own evaluation models and implementation choices, which makes direct comparison of their reported metrics difficult and potentially misleading. We therefore introduce a unified evaluation module, denoted \textit{Evaluator}, that applies the same metric definitions to all translated compiler outputs.

The evaluator focuses on three fidelity components: decoherences fidelity, gates fidelity, and movements fidelity. Their product defines the approximate success probability, which serves as a simple aggregate measure and is broadly consistent with the evaluation style used by the selected compilers. Because the neutral-atom platform can address connectivity constraints through physical shuttling, the number of SWAP gates alone is no longer a sufficiently informative metric of mapping and routing quality \cite{schmid2023hybridcircuitmappingleveraging}. In contrast, the combination of decoherences, gates, and movements fidelity captures both logical gate overhead and the physical cost of shuttling. The following subsections review how these quantities are modeled in HybridMapper, DasAtom, and Enola and highlight the main sources of inconsistency among them.

\subsubsection{HybridMapper}
HybridMapper uses the following definition of approximate success probability \textit{P} proposed in \cite{Schmid_2024}:
\[
 P(C) = exp(-\frac{t_{idle}}{T_{eff}})\prod_{o \in O}\mathcal{F}_o
\]
\[
T_{eff} = \frac{T_1T_2}{T_1 + T_2}, \quad t_{idle} = n \cdot T - \sum_{o \in O}T_o
\]
where $T_1$ and $T_2$ are energy-relaxation and dephasing times, respectively; $T_{eff}$ is the resulting effective coherence time;
$\mathcal{F}_o$ is the fidelity contribution associated with operation $o$; 
$n$ is the number of qubits; $T$ is the total circuit run time; $T_o$ is the duration of operation $o$; and $t_{idle}$ is the total idle time.

This formulation is convenient when each relevant error source can be represented by a scalar fidelity factor in $[0,1]$, as assumed in HybridMapper \cite{schmid2023hybridcircuitmappingleveraging}. However, the set $O$ is not specified explicitly, and the exact formulas used to compute the individual fidelity terms are not provided. In particular, it remains unclear whether SLM $\leftrightarrow$ AOD transitions and cell-to-cell movements are included in $O$. If they are included, their durations would be subtracted from idle time under the above formulation from \cite{Schmid_2024}, which would effectively improve the decoherences term. This treatment differs from DasAtom and Enola, which model movement duration as contributing to overall run time and thus to decoherences.

In addition, HybridMapper reports the negative logarithm of the approximate success probability of the mapped circuit relative to the original circuit. 
This choice may be difficult to interpret because the original circuit is not yet executable on the target hardware and therefore does not admit the same post-compilation fidelity model; for example, no mapping has yet been performed and the routing problem remains unsolved.

\subsubsection{DasAtom}
The next considered compilation toolchain is DasAtom \cite{huang2025dasatomdivideandshuttleatomapproach}. It also relies on approximate success probability \textit{P} as described in \cite{Schmid_2024} but instantiates it in more explicit form:
\[
P(C) = exp(-\frac{t_{idle}}{T_2} ) \times f_{cz}^m \times f_{trans}^s
\]
\[
t_{idle} = n \times T - m \times t_{cz}
\]
\[
T = h \times t_{cz} + s \times t_{trans} + \frac{D}{\upsilon}
\]

Here, $t_{idle}$ is the total idling time summed over all qubits, and $T_2$ is the dephasing time, as in HybridMapper and Enola. The factors $f_{cz}$ and $f_{trans}$ correspond to the fidelity of a CZ gate and an atom transfer, respectively. The quantities $m$ and $s$ are the numbers of CZ gates and atom transfers, $T$ is the total circuit run time, $h$ is the depth of the compiled circuit, $D$ is the sum of the maximal movement distances within each parallel movement group, and $\upsilon$ is the movement speed.

Compared with HybridMapper, DasAtom states more explicitly which quantities are taken into account and how they are used. For example, it does not include one-qubit gates fidelity, as also stated in its paper \cite{huang2025dasatomdivideandshuttleatomapproach}. Also, DasAtom can produce a different number of gates for the same circuit than HybridMapper and Enola \cite{Bachelorarbeit_Emil_Khusainov}. This indicates a possible inconsistency between code generation and metric calculation, which is discussed in more detail in \cref{sec:evaluation}.

Moreover, DasAtom uses $T_2$ directly instead of $T_{eff}$, unlike HybridMapper. This is another source of discrepancy.
On the other hand, DasAtom adds both transfer time and actual movement time to the overall circuit run time and therefore to the decoherences term, in contrast to HybridMapper.

\subsubsection{Enola}
Enola models three main sources of fidelity loss: two-qubit gates, atom transfers between arrays, and qubit decoherences \cite{Enola}. A noticeable difference comparing to DasAtom is the movement-time model. DasAtom assumes a linear dependence of movement time $t$ on distance $d$, as follows: $t = \frac{d}{\upsilon}$, whereas Enola uses a model that also reflects acceleration and deceleration during AOD motion $t = \frac{d}{\upsilon^2}$ \cite{Bluvstein2022-bf}. In our previous study, this component was also found to be sensitive to hard-coded source-code parameters, which led to large metric deviations when Enola was compared directly with other compilers under different hardware settings \cite{Bachelorarbeit_Emil_Khusainov}. The overall formula is computed as follows: 
\[
P(C)  = f_{1q}^{g_1} \cdot f_{cz}^{g_2} \cdot f_{exc}^{|Q|S-2g_2} \cdot f_{trans}^{s} \cdot \prod_{q \in Q} (1 - \frac{T_{q}}{T_2})
\]
where $g_1$ and $g_2$ are the numbers of one-qubit and two-qubit gates, respectively; $Q$ is the set of qubits; $S$ is the number of stages; $|Q|S - 2g_2$ counts qubits affected by the Rydberg laser but not actively participating in a two-qubit gate; $s$ is the total number of atom transfers; $T_2$ is the dephasing time; and the term $T_q$ denotes the idling time of qubit $q$.

A notable difference is the decoherences term. Enola uses $T_2$ instead of $T_{eff}$, which would also account for $T_1$, and it applies a first-order approximation rather than an exponential form. This introduces an additional source of inconsistency when the results are compared directly with those of DasAtom and HybridMapper.

The term $f_{exc}$ models fidelity impact on neighboring qubits. DasAtom does not include this term, and HybridMapper does not define an analogous contribution explicitly, which further complicates direct comparison. Moreover, Enola does not account for one-qubit gates fidelity and sets $f_{1q}=1$ \cite{Tan_2024}. Since DasAtom may produce more one-qubit gates for the same initial circuit and hardware parameters \cite{Bachelorarbeit_Emil_Khusainov}, this difference can also bias the comparison. Similar to DasAtom, Enola includes both inter-array transfers and movement time in the run time contributing to decoherences, in contrast to HybridMapper.

\subsection{Evaluation proposal}
\label{subsec:Evaluator}
To evaluate three selected compilers under aligned assumptions, we define a compiler-agnostic approximation of success probability with explicitly specified fidelity sources. The model is intentionally simple, extensible, and directly computable from RSQASM together with the proposed target hardware description. The purpose of this model is not to reproduce every compiler-specific modeling choice exactly, but to provide a consistent basis for comparison:

\[
 P(C) = \mathcal{F}_{decoherences} \cdot \mathcal{F}_{gates} \cdot \mathcal{F}_{movements}
\]
\[
\mathcal{F}_{decoherences} = exp(-\frac{t_{idle}}{T_{eff}}),
\]
\[
\mathcal{F}_{gates} = \prod_{i = 1}^N f_{g_i},
\]
\[
\mathcal{F}_{movements} = \prod_{i = 1}^s  f_{trans}^{2},
\]
\[
T_{eff} = \frac{T_1T_2}{T_1 + T_2}, \quad t_{idle} = n \cdot T - \sum_{i = 1}^N t_{g_i}
\]
where $N$ denotes the number of executed gates and $s$ is the number of shuttling operations. For each gate $g_i$, the architecture description specifies its average fidelity $f_{g_i}$ and execution time $t_{g_i}$. The parameter $f_{trans}$ is the fidelity contribution of a single $SLM \leftrightarrow AOD$ transfer, $n$ is the number of qubits, and $T$ is the total circuit run time.

For simplicity, we consider three main sources of fidelity loss: decoherences during idling, execution of gates, and physical movements. These sources are among the primary error contributors considered by Enola and DasAtom  \cite{Enola, huang2025dasatomdivideandshuttleatomapproach}. 

\begin{itemize}
  \item \texttt{Decoherences fidelity} is mainly inspired by HybridMapper \cite{schmid2023hybridcircuitmappingleveraging}, because it uses an exponential model and incorporates both hardware decoherence times $T_1$ and $T_2$ through $T_{eff}$ \cite{Schmid_2024}, whereas the other approaches use only $T_2$ or a non-exponential approximation. For compactness, we use the total idle time summed over all qubits, which yields the term $n \cdot T$, where $T$ is the run time of the whole circuit, including gates, movements, and idle periods. The definition of $t_{idle}$ is inspired by DasAtom and Enola in the sense that movements contribute to total run time and therefore also to decoherences, unlike in HybridMapper \cite{schmid2023hybridcircuitmappingleveraging}. The total run time $T$ is computed as the sum of stage execution times, where each stage consists of operations that can be executed in parallel, and the stage duration is determined by the longest operation within that stage. For a movement operation, the time is computed as $2t_{trans} + D / \upsilon$, where $D$ is the Euclidean distance between source cell $s$ and target cell $p$, scaled by the inter-qubit distance specified in the hardware description, i.e., $D = \|s - p\|_2 \cdot size$, and $\upsilon$ is the movement speed in \si{\micro\meter\per\micro\second}.  

  \item \texttt{Gates fidelity} is inspired by all three considered compilers, but reformulated in a consistent way. In particular, both one-qubit and two-qubit gates are included, unlike in Enola and DasAtom, because one-qubit gates can have a noticeable aggregate effect \cite{Bachelorarbeit_Emil_Khusainov}. For simplicity, the model does not include neighbor-qubit excitation effects as in Enola, and it does not reinterpret shuttling as an average gate, since this would introduce inconsistencies between fidelity accounting and idle-time accounting. The resulting model is the product of the average fidelities of all executed native gates, using the values specified in the target hardware description.

  \item \texttt{Movements fidelity} is also inspired by the selected compilers. In the proposed model, the direct fidelity effect of movement is captured through transitions between SLM and AOD. Since each shuttling operation involves two such transitions, the factor $f_{trans}$ appears twice per movement. The indirect impact of shuttling on approximate success probability is already captured by decoherences fidelity, because movement time increases total circuit run time. This separation avoids the inconsistency that arises when movement duration is treated as active gate time and thereby improves decoherences artificially, as in HybridMapper. Consequently, movements fidelity is modeled as the product of two transition-fidelity factors for each of the $s$ movement operations.
  
\end{itemize}

\subsection{Adapter Scripts}
\label{subsec:AdapterScripts}
With the \hyperref[subsec:Evaluator]{Evaluator}, its metric computation methodology, and the common circuit representation RSQASM now defined, the remaining challenge is to translate compiler-specific outputs into RSQASM, i.e., into the unified representation introduced in this work.
This requires exposing and interpreting the intermediate representations of the considered compilers in order to recover information that is not preserved explicitly in their final outputs, such as omitted single-qubit gates, implicit qubit placements, or compiler-specific shuttling constraints. It further requires translating compiler-specific coordinate systems into the evaluator grid.

To demonstrate the feasibility of this process, we provide reference translation tools for the selected compilers, denoted \textit{Adapter Scripts}. Each adapter receives compiler-specific artifacts together with the corresponding architecture description as input and produces (i) a unified architecture file and (ii) an RSQASM representation. These outputs can then be processed by the \textit{Evaluator} under identical assumptions.

For consistency, all input circuits are first transpiled to the same native gate set without optimization, e.g., by using \textit{Qiskit.transpile} with \texttt{optimization\_level = 0}. In this work, we assume the native gate set $\{cz, rx, ry, rz, h, s, t\}$. The role of this preprocessing step is discussed further in Section~\ref{subsec:CompilerIssues}, since it was itself a source of evaluation discrepancies. In addition, each compiler grid is embedded into the top-left corner of the evaluator grid while preserving the inter-qubit-distance invariant. This allows the evaluator to operate directly on grid cells rather than on compiler-specific qubit identifiers.

The evaluator, adapter scripts, and the surrounding automated infrastructure for the selected compilers are available as open-source software on GitHub \footnote{\url{https://github.com/emilkanic0909/UnifiedNeutralAtomToolchainEvaluation}}  and can be used to evaluate additional circuits.

\paragraph{Challenge}
HybridMapper is the simplest case, because its output is already structurally close to RSQASM. Accordingly, the translation mainly involves harmonizing the architecture description and converting the generated output into a runnable RSQASM file.

\paragraph{Approach}
The adapter takes as input the circuit passed to HybridMapper together with the corresponding architecture file. From the number of rows and columns, it derives the evaluator grid dimensions. The parameter \textit{interQubitDistance} is used to define the physical cell size in \si{\micro\meter}. Since the considered version of HybridMapper does not support explicit initial placement, qubits are initially placed on the evaluator grid in row-major order starting from the top-left corner. Gate durations, gate fidelities, shuttling parameters, and decoherence times are then translated lossless into the unified architecture format.

The remaining step is the construction of RSQASM. HybridMapper provides two relevant output formats: a basic format and a more detailed format. Both are close to RSQASM because they are based on a Qiskit-QASM-style representation of mapped circuits. To obtain a runnable RSQASM file from the basic output, it is sufficient to remove register declarations, comments, and include statements, and then prepend the header \texttt{RSQASM 1.0;}. The detailed output additionally contains information about parallel movements and a more precise simulation of AOD movement constraints. However, no dedicated translator for detailed mode is provided here, because the purpose of this work is to expose evaluation discrepancies rather than to compare the degree of detail of compiler-specific low-level movement models.

\subsubsection{Adapter Script of Enola}

\paragraph{Challenge}
Translating Enola artifacts into RSQASM and the unified architecture format requires reconstruction of the semantics of its compiler output.

\paragraph{Approach}
Enola exposes its compilation result as a JSON file containing detailed execution steps, including initialization, different movement types, and applications of Rydberg pulses, i.e., two-qubit gates. In addition, each step records the current grid state, which makes it possible to determine which qubits participate in gates and which atoms are located in SLM and AOD traps. This information is sufficient to construct an RSQASM representation of the mapped circuit. Moreover, it is sufficient to reconstruct omitted single-qubit gates. For this purpose, the instrumented version of Enola in our repository was minimally modified to expose the transpiled circuit used during reconstruction.

The architecture translation is less direct. Enola includes a richer set of hardware constraints than required by the evaluator. Therefore, the current adapter retains only the parameters required by the unified evaluation model. Implementation issues discovered during this process are discussed separately in Section~\ref{subsec:CompilerIssues}.

\subsubsection{Adapter Script of DasAtom}

\paragraph{Challenge}
DasAtom does not expose all information required for reconstruction directly. In its original form, it provides only partial mapping information together with already computed metrics, which is insufficient for a fair comparison.

\paragraph{Approach}
Therefore, the repository used in this work contains an adapted version of DasAtom that additionally exposes the required internal artifacts, including placement graphs, Directed Acyclic Graphs (DAGs), the transpiled circuit, and embeddings.

DasAtom uses a different mapping representation from the other toolchains. In particular, it produces a sequence of embeddings, where each embedding is an array of coordinate pairs and the element at index $i$ specifies the $(x,y)$ position of qubit $i$. For each embedding, DasAtom provides a list of two-qubit gates together with possible parallelization groups. The adapter reconstructs omitted single-qubit gates analogously to the Enola adapter. In addition, DasAtom provides a list of movements required to route qubits from one embedding to the next, again together with possible parallelization information.

However, the provided movement list does not by itself guarantee conflict-free occupation of intermediate cells during execution. For example, a cycle such as $1 \rightarrow 2$, $2 \rightarrow 3$, $3 \rightarrow 1$ cannot be executed directly without temporary relocation. As a conservative resolution, the adapter first moves the qubits one by one into the next row as an intermediate state and then routes them from that row to their target positions while preserving, as far as possible, the parallelization structure defined by the original list. This heuristic is intentionally conservative: it avoids granting DasAtom an artificial advantage through manual optimization while also avoiding the inconsistent omission of required intermediate movements.

Architecture description of DasAtom translates relatively straightforwardly into the proposed format.

\subsubsection{Issues in Existing Compiler Implementations}
\label{subsec:CompilerIssues}

While developing the adapter scripts, several issues in the considered compiler implementations became apparent. These issues are not the main contribution of this work; however, they further motivate the need for a compiler-agnostic post-compilation representation and an independent evaluation workflow.

\paragraph{Inconsistent input transpilation}
An important inconsistency was identified in the input transpilation stage. DasAtom invokes \textit{qiskit.transpile} with optimization\_level 0, whereas Enola omits this parameter. According to the IBM Qiskit documentation \cite{QisKit_doc}, omitting this parameter results in optimization\_level = 2; in older Qiskit versions such as the one used by Enola, omitting the parameter results in level 1 \cite{qiskit_1_1_0_commit}. This directly changes the number of gates in the compiled input and therefore makes the comparison of mapping and routing quality inconsistent.

This issue is substantial because mapping-oriented compilers should not be compared on effectively different input circuits. HybridMapper does not include its own transpilation stage and instead operates directly on circuits already expressed in the target architecture gate set. To ensure consistency across all considered toolchains, we therefore introduce the same initial transpilation step for all inputs, using the same gate set and optimization\_level = 0 before passing the circuit to the respective adapted compiler version. In this way, every compiler receives the same native-gate input into its mapping, routing, and scheduling stages.

\paragraph{Incorrect interpretation of two-qubit operations in Enola}
During translation into RSQASM, an additional implementation bug was identified in Enola. Specifically, Enola treated all two-qubit operations in the input Qiskit QASM circuit as CZ gates; for example, a \texttt{barrier} operation in QASM was incorrectly interpreted as a CZ gate and routed accordingly \cite{Enola_commit}.

\begin{lstlisting}[language=rsqasm,
caption={Example QASM illustrating misinterpreted two-qubit operations.},
label={lst:qasm_Enola_Issue}]
cz q[1], q[2];
barrier q[1], q[2];
\end{lstlisting}

This issue became apparent during the RSQASM translation process because the adapter script performs consistency checks on trap occupancy and qubit positions. After translation, an attempt to move a qubit from an unoccupied trap resulted in an error, which led to the identification of the underlying bug in Enola. It was subsequently fixed in the instrumented version of Enola, together with the parameter-setting inconsistencies discussed in \cite{Bachelorarbeit_Emil_Khusainov}.

\paragraph{Architecture-related inconsistencies in Enola}
A further issue concerns the interpretation of architecture parameters in Enola. During our investigation, several parameter combinations led to implementation problems that would require substantial additional engineering effort to resolve completely. In particular, Enola may ignore the nominal grid dimensions and produce coordinates beyond the declared architecture size. Accordingly, the adapter determines the evaluator grid size from the maximum coordinate actually used in the compiled output. This choice is pragmatic and sufficient for the evaluation methodology proposed in this work.

%% file: sections/evaluation.tex
Having adapted the compiler sources, the \hyperref[subsec:AdapterScripts]{Adapter Scripts}, and the \hyperref[subsec:Evaluator]{Evaluator}, we can perform a controlled comparative study of the selected toolchains. The purpose of this evaluation is not to claim a definitive or exhaustive benchmarking solution. Rather, it serves as a proof of concept demonstrating that differences in metric definitions, hardware abstractions, and post-compilation representations can affect the reported results and therefore motivate the need for a more standardized evaluation methodology.

\textbf{Common Hardware Assumptions.} To this end, we fix a common set of hardware parameters for both the considered compilers and the \hyperref[subsec:Evaluator]{Evaluator}. For the coherence times, we set $T_2 = \SI{1.5}{\second}$, consistent with \cite{schmid2023hybridcircuitmappingleveraging, Wintersperger_2023, Graham2022-pl, Bluvstein2022-bf, huang2025dasatomdivideandshuttleatomapproach, Enola}, and $T_1 = \SI{100}{\second}$, following \cite{schmid2023hybridcircuitmappingleveraging}.
Although longer coherence times, e.g., up to \SI{40}{\second} for nuclear spins, have been reported in \cite{Wintersperger_2023, Barnes2022-gh} and can influence the resulting fidelity metrics, the present study uses the more common setting summarized in \cref{tab:hardware}.

For gate fidelities, we use 0.9996 for two-qubit gates, consistent with values commonly used in compiler case studies and compatible with \cite{Wintersperger_2023, Graham2022-pl}, and 0.9999 for single-qubit gates, following \cite{schmid2023hybridcircuitmappingleveraging}. For gate durations, we set \SI{0.2}{\micro\second} for two-qubit gates, corresponding to Rydberg interactions, and \SI{2}{\micro\second} for single-qubit gates, corresponding to Raman interactions, in line with the order of magnitude reported in \cite{Wintersperger_2023, Bluvstein2022-bf}. The AOD activation/deactivation time is set to \SI{20}{\micro\second}, as in \cite{huang2025dasatomdivideandshuttleatomapproach}. The transfer fidelity is set to 0.9999, following HybridMapper and remaining consistent with the non-negligible transfer effects discussed in \cite{Wintersperger_2023}. Finally, the movement speed is set to $0.55\,\si{\micro\meter\per\micro\second}$, as in DasAtom and Enola.

\begin{table}[htpb]
  \caption[Hardware]{Hardware parameters used in the case study.}\label{tab:hardware}
  \centering
  \begin{tabular}{l l}
    \toprule
      Parameter & Value \\
    \midrule
      Evaluator grid side length & 50 \\
      Evaluator inter-qubit distance \si{\micro\meter} & 1 \\
      Two-qubit gate time \si{\micro\second} & 0.2 \\
      One-qubit gate time \si{\micro\second} & 2 \\
      Two-qubit gate fidelity & 0.9996 \\
      One-qubit gate fidelity & 0.9999 \\
      Coherence Time $T_1$ \si{\micro\second} & 100000000 \\
      Coherence Time $T_2$ \si{\micro\second} & 1500000 \\
      AOD activation/deactivation time \si{\micro\second} & 20 \\
      $AOD \leftrightarrow SLM$ transfer fidelity  & 0.9999 \\
      Movement speed \si{\micro\meter\per\micro\second} & 0.55 \\
    \bottomrule
  \end{tabular}
\end{table}

This section focuses primarily on the claims reported in DasAtom, because its paper contains broad direct comparisons against other neutral-atom compilers, including Enola, on a relatively rich benchmark set \cite{huang2025dasatomdivideandshuttleatomapproach}. 
Rather, it provides a particularly informative case study for examining how strongly reported performance gaps may depend on the underlying evaluation methodology. By contrast, Enola reports comparisons mainly against OLSQ\_DPQA, the predecessor of itself \cite{Enola}, while HybridMapper primarily compares different variants of itself \cite{schmid2023hybridcircuitmappingleveraging}.

For the considered case studies, we report results for DasAtom, Enola, and the three HybridMapper variants corresponding to the SWAP-based, move-based, and hybrid routing modes (see \cref{sec:prelim}). As discussed in \cref{sec:methods}, Enola uses a substantially richer architecture description than the other compilers, which makes exact alignment of some geometric parameters difficult. \textbf{Issue (geometry mismatch).} DasAtom reports an inter-qubit distance of \SI{3}{\micro\meter} for itself and \SI{15}{\micro\meter} for Enola \cite{huang2025dasatomdivideandshuttleatomapproach}, which, together with an interaction radius of \SI{6}{\micro\meter}, yields a different effective ratio between interaction range and qubit spacing. Under these parameters, the interaction radius is smaller than the inter-qubit distance in the Enola configuration, meaning that two-qubit gates cannot be executed directly. As a result, the configurations are not geometrically comparable, which prevents a consistent and fair comparison. \textbf{Result.} in the present study, the evaluator instead assumes a unit cell spacing of \SI{1}{\micro\meter} in the horizontal and vertical directions, i.e., using orthogonal rather than diagonal adjacency, and applies this abstraction consistently across all translated outputs. Moreover, for all compilers, we attempt to configure the architecture such that the inter-qubit distance corresponds to \SI{1}{\micro\meter} whenever this could be controlled through the compiler interface and architecture description. The resulting comparison should therefore be understood as an evaluation under a unified post-compilation abstraction, rather than as an attempt to reproduce each compiler's full native hardware model exactly.

\subsection{QFT30: Re-evaluating the Reported 415.8x Gap at Interaction Radius \SI{6}{\micro\meter}}
DasAtom reports a 415.8x fidelity improvement over Enola on the Quantum Fourier Transform (QFT) benchmark with 30 qubits (QFT30) and suggests that the improvement is exponential \cite{huang2025dasatomdivideandshuttleatomapproach}. This claim is therefore a natural starting point for re-evaluation. In the reported DasAtom setup, the interaction radius is set to \SI{6}{\micro\meter}. Since the hardware parameters listed in \cref{tab:hardware} are already close to those used in DasAtom, we keep the common hardware setting fixed throughout this section and vary only the interaction radius, since interaction radius is crucial for mapping and routing.
\begin{table}[htpb]
  \caption[QFT30, interaction radius 6\,\textmu m. Results in (\%).]{QFT30, interaction radius 6\,\textmu m. Results in (\%).}\label{tab:qft30_R6}
  \centering
  \begin{tabular}{l c c c c}
    \toprule
      Compiler & $\mathcal{F}_{decoh}$ & $\mathcal{F}_{gates}$ & $\mathcal{F}_{moves}$ & ASP \\
    \midrule
      DasAtom & 88.01 & 50.53 & 100 & 44.47\\
      Enola & 15.58 & 50.53 & 69.37 & 5.46 \\
      HybridMapper SWAP & 84.09 & 35.63 & 100 & 29.96 \\
      HybridMapper MOVE & 83.41 & 50.53 & 99.02 & 41.73 \\
      HybridMapper Hybrid & 83.41 & 50.53 & 99.02 & 41.73 \\
    \bottomrule
  \end{tabular}
  \parbox{0.9\linewidth}{\footnotesize
  ASP = Approximate Success Probability.
  }
\end{table}  
 
\cref{tab:qft30_R6} leads to several important observations. The previously reported 415.8x advantage of DasAtom over Enola does not reappear under the unified evaluation pipeline. Instead, the gap in the approximate success probability (ASP) is approximately \(44.47 / 5.46 \approx 8.1\), which is still substantial but far smaller than the originally reported value. 

\textbf{Analysis.} DasAtom performs no shuttling operations in this setting, as the movement fidelity remains at 100\%. A plausible explanation is that, with interaction radius \SI{6}{\micro\meter} and evaluator inter-qubit distance \SI{1}{\micro\meter}, the qubits can be arranged so that all required interactions are achievable without physical movements. 

Enola loses a large fraction of fidelity through decoherences, which indicates a high accumulated idle time. This effect is examined in more detail below. Finally, the move-based and hybrid variants of HybridMapper achieve ASP values close to that of DasAtom. For this circuit and under aligned assumptions, this suggests that routing strategy alone is unlikely to explain the extreme and possible exponential gap reported in prior work.

One possible reason for the absence of movements in DasAtom is the ratio between the interaction radius and the inter-qubit distance. In DasAtom, the reported setting uses an interaction radius of \SI{6}{\micro\meter} together with an inter-qubit distance of \SI{3}{\micro\meter}, i.e., an effective ratio of 2. In contrast, for Enola the inter-qubit distance is set to \SI{15}{\micro\meter}, which does not preserve this effective ratio and may therefore lead to substantially more shuttling operations. 

\textbf{Result.} Under the evaluator grid with an inter-qubit distance of \SI{1}{\micro\meter}, the corresponding interaction radius would thus be \SI{2}{\micro\meter} rather than \SI{6}{\micro\meter}. This motivates an additional experiment with ratio 2 and therefore interaction radius \SI{2}{\micro\meter}. For Enola, however, the inter-qubit distance cannot be controlled conveniently through its architecture description, as discussed in \cref{sec:methods}. Therefore, an exact proportional alignment of this ratio cannot be enforced.

\subsection{QFT30: Re-evaluating the Reported 415.8x Gap at Interaction Radius \SI{2}{\micro\meter}}
As discussed in the previous subsection, the interaction radius \SI{6}{\micro\meter} setting does not preserve the effective ratio between interaction radius and inter-qubit distance reported for DasAtom. To examine whether this geometric mismatch contributes to the observed discrepancy, we now reduce the interaction radius to \SI{2}{\micro\meter}. This setting is intended to reduce the mismatch between interaction radius and inter-qubit distance ratio and therefore to provide a more comparable geometry across the translated compiler outputs, while still remaining within the simplified evaluator abstraction.
\begin{table}[htpb]
  \caption[QFT30, interaction radius 2\,\textmu m. Results in (\%).]{QFT30, interaction radius 2\,\textmu m. Results in (\%).}
  \label{tab:qft30_R2}
  \centering
  \begin{tabular}{l c c c c}
    \toprule
      Compiler & $\mathcal{F}_{decoh}$ & $\mathcal{F}_{gates}$ & $\mathcal{F}_{moves}$ & ASP \\
    \midrule
      DasAtom & 59.39 & 50.53 & 88.58 & 26.58\\
      Enola & 15.57 & 50.53 & 69.37 & 5.46 \\
      HybridMapper SWAP & 58.31 & 2.46 & 100 & 1.43 \\
      HybridMapper MOVE & 70.67 & 50.53 & 95.33 & 34.04 \\
      HybridMapper Hybrid & 70.05 & 50.53 & 95.19 & 33.69 \\
    \bottomrule
  \end{tabular}
  \parbox{0.9\linewidth}{\footnotesize
  ASP = Approximate Success Probability.
  }
\end{table} 

The results in \cref{tab:qft30_R2} make the discrepancy with the originally reported 415.8x gap even more apparent. DasAtom still outperforms Enola, but the gap remains far below the previously claimed value and now is approximately \(26.58 / 5.46 \approx 4.86\). This result suggests that aligned assumptions and a unified post-compilation abstraction can materially affect the comparison.

\textbf{Analysis.} A noteworthy observation is that, in the present setup, Enola produces exactly the same result for interaction radius of 6 and 2. This may indicate that the chosen Enola architecture parameters do not materially change the relevant spacing constraints under the translated model, or that additional architecture parameters influence the effective behavior in a way not fully exposed through the available interface.

The move-based and hybrid modes of HybridMapper now outperform DasAtom in terms of ASP. This further supports the view that very large reported gaps may arise not only from compiler strategy itself, but also from evaluation assumptions. At the same time, this result should be interpreted cautiously: HybridMapper solves routing in place and does not use DasAtom's subcircuit decomposition strategy, so the relative ranking may still depend strongly on the circuit family under consideration.

Finally, the poor ASP of HybridMapper SWAP is expected. A SWAP-based approach is disadvantaged when compared directly against shuttling-based under a metric model that was mainly considered for exploring shuttling-based approaches. The choice of fidelity parameters for evaluation between SWAP-based and move-based is critical.

\subsection{Explaining Low Decoherences Fidelity of Enola}
To localize the source of Enola's low decoherences fidelity, we inspect the generated RSQASM for QFT30 at interaction radius 2. This reveals several patterns of redundant or overly detailed movement instructions. The following analysis should not be interpreted as a complete diagnosis of Enola's internal routing strategy. Rather, it illustrates how differences in the abstraction level of the exposed compiler output can already influence the metrics produced by the evaluator.

Consider the following fragment:
\begin{lstlisting}[language=rsqasm,
caption={Example of immediately reversed movement.},
label={lst:rsqasm_reverse}]
cz q[1034], q[1033];
move q[1034], q[1028];
move q[1028], q[1034];
h q[1034];
\end{lstlisting} 

As noted earlier, RSQASM indices denote cell positions on the grid rather than qubit identifiers. Consequently, moving from cell 1034 to 1028 and then immediately back to 1034 does not change the effective grid state at the RSQASM level.

A second pattern is illustrated below:
\begin{lstlisting}[language=rsqasm,
caption={Example of movement path with an unnecessary intermediate cell.},
label={lst:rsqasm_intermediate}]
cz q[1034], q[1033];
move q[1034], q[865];
move q[865], q[1029];
move q[1201], q[1034];
h q[1028];
\end{lstlisting} 

Here, the movement path $1034 \rightarrow 865 \rightarrow 1029$ can be collapsed into a single effective movement $1034 \rightarrow 1029$, provided that the intermediate cell serves only as a low-level routing artifact and does not change the logical state of the mapped circuit at the abstraction level used by the evaluator.

Such patterns can also appear in nested form:
\begin{lstlisting}[language=rsqasm,
caption={Nested redundant movement pattern.},
label={lst:rsqasm_nested}]
cz q[1029], q[1028];cz q[1034], q[1033];
move q[1034], q[1201];
move q[1029], q[865];
move q[865], q[1029];
move q[1201], q[1034];
h q[1029];
\end{lstlisting}

\textbf{Result.} At the RSQASM level, all four movements in \cref{lst:rsqasm_nested} can be removed without changing the effective grid state before the subsequent logical operation. This does not necessarily mean that these movements are unnecessary in Enola's internal low-level model. They may still be required to satisfy architectural constraints such as non-crossing AOD lines or ordering constraints between AOD channels. However, the goal of the present framework is to compare compilers at a shared post-compilation abstraction level. Since HybridMapper and DasAtom already expose more abstract movement descriptions, retaining such low-level intermediate movements in Enola would penalize it unfairly under the unified evaluator.

\subsection{QFT30, Interaction Radius \SI{2}{\micro\meter}: Collapsing of RSQASM-Redundant Movements}
As discussed in the previous subsection, the translated Enola output contains movement patterns that are redundant at the RSQASM abstraction level. We therefore collapse them. This should not be interpreted as an optimization of Enola's routing strategy. Rather, it is a normalization step intended to align the abstraction level of the translated outputs across all considered toolchains.

For simplicity, only syntactically obvious redundant shuttling patterns are removed. The goal is not to obtain the best possible Enola schedule, but to estimate how much fidelity can be lost solely because one compiler exposes a lower-level movement representation than the others. After this normalization step, the resulting sequence is sufficient to estimate the change in movement count and traveled distance, even though it is not reintroduced as a new executable RSQASM schedule.

For QFT30 at radius \SI{2}{\micro\meter}, the following differences are obtained:
\begin{table}[htpb]
  \caption[QFT30, radius 2\,\textmu m: effect of collapsing redundant Enola movements.]{QFT30, radius 2\,\textmu m: effect of collapsing redundant Enola movements.}
  \label{tab:qft30_R2_diss}
  \centering
  \begin{tabular}{l c}
    \toprule
      Metric & Value \\
    \midrule
    Saved Distance (cell units) & 6003.69  \\
    Movement count before collapsing & 1828  \\
    Movement count after collapsing & 937  \\
    \bottomrule
  \end{tabular}
\end{table}

Using the evaluator model from \cref{sec:methods} and the hardware parameters from \cref{tab:hardware}, the reduction in distance-dependent movement time is:
\[
\Delta T_{move} = \frac{6003.69 \cdot 1}{0.55} = \SI{10915.8}{\micro\second}.
\] Since this run time reduction affects all 30 qubits, the summed idle-time reduction is
\[
\Delta t_{idle} = 30 \cdot \Delta T_{move} = \SI{327474}{\micro\second}.
\] 
With \(t_{idle}^{old} = \SI{2747600}{\micro\second}\), this yields
\[
t_{idle}^{new} = t_{idle}^{old} - \Delta t_{idle}
= 2747600 - 327474
= \SI{2420126.37}{\micro\second}.
\]
 Then the updated movements fidelity is
 \[
\mathcal{F}_{moves} = f_{trans}^{2 \cdot N_{moves}} = 0.9999^{2 \cdot 937} = 0.8291,
\]
i.e., \(82.91\%\). The updated decoherences fidelity is
\[
\mathcal{F}_{decoh} = \exp\left(-\frac{t_{idle}^{new}}{T_{eff}}\right)
= \exp\left(-\frac{2420126.37}{1477832.51}\right)
= 0.1944,
\]
i.e., \(19.44\%\).
\begin{table}[htpb]
  \caption[QFT30 Radius 2\,\textmu m Enola Collapsed. Results in (\%).]{QFT30 Radius 2\,\textmu m Enola Collapsed. Results in (\%).}\label{tab:qft30_R2_Enola_Opt}
  \centering
  \begin{tabular}{l c c c c}
    \toprule
      Compiler & $\mathcal{F}_{decoh}$ & $\mathcal{F}_{gates}$ & $\mathcal{F}_{moves}$ & ASP \\
    \midrule
      DasAtom & 59.39 & 50.53 & 88.58 & 26.58\\
      Enola & 15.57 & 50.53 & 69.37 & 5.46 \\
      Enola (collapsed) & 19.44 & 50.53 & 82.91 & 8.14 \\
      HybridMapper SWAP & 58.31 & 2.46 & 100 & 1.43 \\
      HybridMapper MOVE & 70.67 & 50.53 & 95.33 & 34.04 \\
      HybridMapper Hybrid & 70.05 & 50.53 & 95.19 & 33.69 \\
    \bottomrule
  \end{tabular}
  \parbox{0.9\linewidth}{\footnotesize
  ASP = Approximate Success Probability.
  }
\end{table} 

The resulting values are summarized in \cref{tab:qft30_R2_Enola_Opt}. Enola's fidelity improves after collapsing RSQASM-redundant movements, and the resulting ASP gap decreases to approximately \(26.58 / 8.14 \approx 3.26\), far below the originally reported factor of 415.8. This supports the broader claim of the paper: even under a deliberately simplified  evaluator, abstraction-level inconsistencies can materially affect the reported metrics. At the same time, the remaining performance gap should not be overinterpreted. It may still depend on implementation details that are not fully exposed through Enola's architecture interface. In particular, the translated Enola outputs tend to place qubits relatively far apart before interactions, but the exact reason for this behavior cannot be determined with confidence from the available artifacts alone. Therefore, the present result should be understood primarily as evidence for the sensitivity of compiler comparisons to evaluation choices, rather than as a complete or final diagnosis of Enola's routing strategy.

\subsection{Additional Benchmark Results}

\begin{table}[htpb]
  \caption[ASP on additional benchmarks at interaction radius 2\,\textmu m. Results in (\%).]{ASP on additional benchmarks at interaction radius 2\,\textmu m. Results in (\%).}\label{tab:bs30_R2}
  \centering
  \begin{tabular}{l c c c c}
    \toprule
      Compiler & BV$_{30}$ & GHZ$_{30}$ & DJ$_{30}$ & QAOA$_{30}$ \\
    \midrule
      DasAtom & 97.23* & 98.02 & 90.8 & 33.29\\
      Enola & 95.99 & 91.63 & 90.07 & 6.97 \\
      HybridMapper SWAP & N/A$^{\dagger}$ & N/A$^{\dagger}$ & 94.83 & N/A$^{\dagger}$ \\
      HybridMapper MOVE & N/A$^{\dagger}$ & N/A$^{\dagger}$ & 95.41 & N/A$^{\dagger}$ \\
      HybridMapper Hybrid & N/A$^{\dagger}$ & N/A$^{\dagger}$ & 95.41 & N/A$^{\dagger}$ \\
    \bottomrule
  \end{tabular}
  \parbox{0.9\linewidth}{\footnotesize
  ASP = Approximate Success Probability.
  * Approximate result obtained in a semi-automated workflow, since the compiler behavior for independent wire of single-qubit gates is not fully exposed.
  $\dagger$ For these cases, the HybridMapper output exhibited apparent mapping/output failures, including inconsistent gate counts and premature termination of the compiled artifact. The reported values should therefore be interpreted with caution.
  }
\end{table}  

While the main discussion in this section focuses on QFT30, motivated by the explicit cross-compiler comparisons reported in DasAtom, it is also useful to broaden the evaluation to additional benchmark circuits. Table~\ref{tab:bs30_R2} therefore reports ASP values for BV$_{30}$, GHZ$_{30}$, DJ$_{30}$, and QAOA$_{30}$ under the same unified methodology. Beyond complementing the QFT30 case study, such reference points can support future comparative studies and help reveal further benchmarking inconsistencies that may remain hidden when compilers are evaluated in isolation.

%% file: sections/relevant_work.tex
Several compiler frameworks for neutral-atom quantum computing have been proposed in recent years. Some of them focus primarily on mapping and routing, while others also incorporate scheduling. As the field is still emerging, a variety of compilation strategies continue to be explored and refined.

Consequently, inconsistencies in benchmarking methodologies and metric definitions have already been discussed for quantum computing as a whole \cite{lall2025reviewcollectionmetricsbenchmarks}. Recent efforts such as \textit{Metriq} aim to address related issues by providing a benchmarking infrastructure for cross-hardware evaluation \cite{cosentino2026metriqcollaborativeplatformbenchmarking}. 

A closely related precursor to this work is the study in \cite{Bachelorarbeit_Emil_Khusainov}, which analyzed inconsistencies in the calculation of Approximate Success Probability for neutral-atom architectures. That study considered the HybridMapper, Enola, and DasAtom toolchains and showed that differences in calculation models, as well as minor fixes in compiler source code, can significantly affect the reported results. To the best of our knowledge, prior work on neutral-atom compilers still lacks a reproducible, compiler-agnostic framework that provides both a unified post-compilation representation and a consistent evaluation methodology.

%% file: sections/conclusion.tex
In this work, we identified inconsistencies in the evaluation of fidelity-related metrics across different neutral-atom quantum compilers. To address this issue, we introduced a hardware-agnostic benchmarking framework with standardized metric evaluation procedures. As a demonstration of feasibility, we extended several existing compilers to export compiled circuits in the proposed format and corrected several implementation issues that had led to incorrect metric computations. The results show that applying a unified evaluation methodology can substantially change the reported outcomes, thereby underscoring the importance of standardized benchmarking practices. In this work, we do not claim that the proposed evaluation methodology is the definitive one. Rather, our goal is to demonstrate how reported results can differ substantially when different evaluation assumptions are used.
Future work includes extending the set of supported compilers, either through native integration with the framework or through transpilers that convert compiler outputs into the proposed representation, as demonstrated in our approach. In addition, the evaluation model can be further refined to incorporate additional error sources.

%% file: references.bib
@misc{schmid2023hybridcircuitmappingleveraging,
      title={Hybrid Circuit Mapping: Leveraging the Full Spectrum of Computational Capabilities of Neutral Atom Quantum Computers}, 
      author={Ludwig Schmid and Sunghye Park and Seokhyeong Kang and Robert Wille},
      year={2023},
      eprint={2311.14164},
      archivePrefix={arXiv},
      primaryClass={quant-ph},
      url={https://arxiv.org/abs/2311.14164}, 
}

@article{Schmid_2024,
   title={Computational capabilities and compiler development for neutral atom quantum processors—connecting tool developers and hardware experts},
   volume={9},
   ISSN={2058-9565},
   url={http://dx.doi.org/10.1088/2058-9565/ad33ac},
   DOI={10.1088/2058-9565/ad33ac},
   number={3},
   journal={Quantum Science and Technology},
   publisher={IOP Publishing},
   author={Schmid, Ludwig and Locher, David F and Rispler, Manuel and Blatt, Sebastian and Zeiher, Johannes and Müller, Markus and Wille, Robert},
   year={2024},
   month=apr, pages={033001} }

@misc{huang2025dasatomdivideandshuttleatomapproach,
      title={DasAtom: A Divide-and-Shuttle Atom Approach to Quantum Circuit Transformation}, 
      author={Yunqi Huang and Dingchao Gao and Shenggang Ying and Sanjiang Li},
      year={2025},
      eprint={2409.03185},
      archivePrefix={arXiv},
      primaryClass={quant-ph},
      url={https://arxiv.org/abs/2409.03185}, 
}

@misc{Bachelorarbeit_Emil_Khusainov,
	author = {Khusainov, Emil},
	title = {Tooling and Benchmarking of a Hardware-Agnostic Compilation Toolchain For Neutral-Atom Quantum Computers},
	year = {2025},
    school = {Technische Universität München},
    language = {en},
    type = {Bachelor's thesis}, 
    url = {https://mediatum.ub.tum.de/node?id=1788052&change_language=en},
}

@inbook{Enola,
author = {Tan, Daniel Bochen and Lin, Wan-Hsuan and Cong, Jason},
title = {Compilation for Dynamically Field-Programmable Qubit Arrays with Efficient and Provably Near-Optimal Scheduling},
year = {2025},
isbn = {9798400706356},
publisher = {Association for Computing Machinery},
address = {New York, NY, USA},
url = {https://doi.org/10.1145/3658617.3697778},
booktitle = {Proceedings of the 30th Asia and South Pacific Design Automation Conference},
pages = {921–929},
numpages = {9}
}

@ARTICLE{Bluvstein2022-bf,
  title    = "A quantum processor based on coherent transport of entangled atom
              arrays",
  author   = "Bluvstein, Dolev and Levine, Harry and Semeghini, Giulia and
              Wang, Tout T and Ebadi, Sepehr and Kalinowski, Marcin and
              Keesling, Alexander and Maskara, Nishad and Pichler, Hannes and
              Greiner, Markus and Vuleti{\'c}, Vladan and Lukin, Mikhail D",
  journal  = "Nature",
  volume   =  604,
  number   =  7906,
  pages    = "451--456",
  month    =  apr,
  year     =  2022
}

@article{Tan_2024,
   title={Compiling Quantum Circuits for Dynamically Field-Programmable Neutral Atoms Array Processors},
   volume={8},
   ISSN={2521-327X},
   url={http://dx.doi.org/10.22331/q-2024-03-14-1281},
   DOI={10.22331/q-2024-03-14-1281},
   journal={Quantum},
   publisher={Verein zur Forderung des Open Access Publizierens in den Quantenwissenschaften},
   author={Tan, Daniel Bochen and Bluvstein, Dolev and Lukin, Mikhail D. and Cong, Jason},
   year={2024},
   month=mar, pages={1281} }

@manual{QisKit_doc,
  title        = {IBM Qiskit Compiler Transpile Documentation},
  organization = {IBM Qiskit},
  year         = {2026},
  url          ={https://quantum.cloud.ibm.com/docs/en/api/qiskit/dev/compiler},
  note         = {Accessed: 2026-03-03 for version 2.4.0}
}

@misc{qiskit_1_1_0_commit,
  author = {{Qiskit Development Team}},
  title  = {qiskit/compiler/transpiler.py line 224 (state by last commit 7d29dc1b33ab0229bc77ab773b8d2f7fd2589552) for release 1.1.0},
  year   = {2026},
  url    = {https://github.com/Qiskit/qiskit/commit/7d29dc1b33ab0229bc77ab773b8d2f7fd2589552},
  note   = {Accessed: 2026-03-03}
}

@misc{Enola_commit,
  author = {{UCLA-VAST Development Team}},
  title  = {Enola/run\_qasm.py line 25 
            (state at commit 2944dbf4e163e8d2eeeec607add0d9139edce689)},
  year   = {2026},
  url    = {https://github.com/UCLA-VAST/Enola/commit/2944dbf4e163e8d2eeeec607add0d9139edce689},
  note   = {Accessed: 2026-03-05}
}

@article{Wintersperger_2023,
   title={Neutral atom quantum computing hardware: performance and end-user perspective},
   volume={10},
   ISSN={2196-0763},
   url={http://dx.doi.org/10.1140/epjqt/s40507-023-00190-1},
   DOI={10.1140/epjqt/s40507-023-00190-1},
   number={1},
   journal={EPJ Quantum Technology},
   publisher={Springer Science and Business Media LLC},
   author={Wintersperger, Karen and Dommert, Florian and Ehmer, Thomas and Hoursanov, Andrey and Klepsch, Johannes and Mauerer, Wolfgang and Reuber, Georg and Strohm, Thomas and Yin, Ming and Luber, Sebastian},
   year={2023},
   month=aug }

@ARTICLE{Graham2022-pl,
  title    = "Multi-qubit entanglement and algorithms on a neutral-atom quantum
              computer",
  author   = "Graham, T M and Song, Y and Scott, J and Poole, C and Phuttitarn,
              L and Jooya, K and Eichler, P and Jiang, X and Marra, A and
              Grinkemeyer, B and Kwon, M and Ebert, M and Cherek, J and
              Lichtman, M T and Gillette, M and Gilbert, J and Bowman, D and
              Ballance, T and Campbell, C and Dahl, E D and Crawford, O and
              Blunt, N S and Rogers, B and Noel, T and Saffman, M",
  journal  = "Nature",
  volume   =  604,
  number   =  7906,
  pages    = "457--462",
  month    =  apr,
  year     =  2022
}

@ARTICLE{Barnes2022-gh,
  title    = "Assembly and coherent control of a register of nuclear spin
              qubits",
  author   = "Barnes, Katrina and Battaglino, Peter and Bloom, Benjamin J and
              Cassella, Kayleigh and Coxe, Robin and Crisosto, Nicole and King,
              Jonathan P and Kondov, Stanimir S and Kotru, Krish and Larsen,
              Stuart C and Lauigan, Joseph and Lester, Brian J and McDonald,
              Mickey and Megidish, Eli and Narayanaswami, Sandeep and
              Nishiguchi, Ciro and Notermans, Remy and Peng, Lucas S and Ryou,
              Albert and Wu, Tsung-Yao and Yarwood, Michael",
  journal  = "Nature Communications",
  volume   =  13,
  number   =  1,
  pages    = "2779",
  month    =  may,
  year     =  2022
}

@misc{lall2025reviewcollectionmetricsbenchmarks,
      title={A Review and Collection of Metrics and Benchmarks for Quantum Computers: definitions, methodologies and software}, 
      author={Deep Lall and Abhishek Agarwal and Weixi Zhang and Lachlan Lindoy and Tobias Lindström and Stephanie Webster and Simon Hall and Nicholas Chancellor and Petros Wallden and Raul Garcia-Patron and Elham Kashefi and Viv Kendon and Jonathan Pritchard and Alessandro Rossi and Animesh Datta and Theodoros Kapourniotis and Konstantinos Georgopoulos and Ivan Rungger},
      year={2025},
      eprint={2502.06717},
      archivePrefix={arXiv},
      primaryClass={quant-ph},
      url={https://arxiv.org/abs/2502.06717}, 
}

@misc{cosentino2026metriqcollaborativeplatformbenchmarking,
      title={Metriq: A Collaborative Platform for Benchmarking Quantum Computers}, 
      author={Alessandro Cosentino and Changhao Li and Vincent Russo and Bradley A. Chase and Tom Lubinski and Siyuan Niu and Neer Patel and Nathan Shammah and William J. Zeng},
      year={2026},
      eprint={2603.08680},
      archivePrefix={arXiv},
      primaryClass={quant-ph},
      url={https://arxiv.org/abs/2603.08680}, 
}

@article{Henriet2020quantumcomputing,
  doi = {10.22331/q-2020-09-21-327},
  url = {https://doi.org/10.22331/q-2020-09-21-327},
  title = {Quantum computing with neutral atoms},
  author = {Henriet, Lo{\"{i}}c and Beguin, Lucas and Signoles, Adrien and Lahaye, Thierry and Browaeys, Antoine and Reymond, Georges-Olivier and Jurczak, Christophe},
  journal = {{Quantum}},
  issn = {2521-327X},
  publisher = {{Verein zur F{\"{o}}rderung des Open Access Publizierens in den Quantenwissenschaften}},
  volume = {4},
  pages = {327},
  month = sep,
  year = {2020}
}

@misc{wang2024atomiquequantumcompilerreconfigurable,
      title={Atomique: A Quantum Compiler for Reconfigurable Neutral Atom Arrays}, 
      author={Hanrui Wang and Pengyu Liu and Daniel Bochen Tan and Yilian Liu and Jiaqi Gu and David Z. Pan and Jason Cong and Umut A. Acar and Song Han},
      year={2024},
      eprint={2311.15123},
      archivePrefix={arXiv},
      primaryClass={quant-ph},
      url={https://arxiv.org/abs/2311.15123}, 
}
